\documentclass[a4paper]{article}
\usepackage{fullpage}
\usepackage[utf8]{inputenc}
\usepackage[english]{babel}
\usepackage{mathtools}
\usepackage{enumitem}

\usepackage{graphicx}
\usepackage{amssymb}
\usepackage{amsmath}

\newtheorem{lemma}{Lemma}

\title{A PTAS for $k$-hop MST on the Euclidean plane: Improving Dependency on $k$}

\author{
  Jittat Fakcharoenphol\thanks{
    Email: jittat@gmail.com,
    Department of Computer Engineering, Kasetsart University, Bangkok, Thailand.
    Supported by the Thailand Research Fund, Grant RSA-6180074.}
  \and
  Nonthaphat Wongwattanakij\thanks{
    Email: nonthaphat.wo@ku.th,
    Department of Computer Engineering, Kasetsart University, Bangkok, Thailand.
    Supported by Graduate Student Scholarship, Faculty of Engineering, Kasetsart
University.}
}

\begin{document}
\maketitle

\begin{abstract}
For any $\epsilon>0$, Laue and Matijevi\'{c}~[CCCG'07, IPL'08] give a PTAS for finding a $(1+\epsilon)$-approximate solution to the $k$-hop MST problem in the Euclidean plane that runs in time $(n/\epsilon)^{O(k/\epsilon)}$.  
In this paper, we present an algorithm that runs in time 
$(n/\epsilon)^{O(\log k \cdot(1/\epsilon)^2\cdot\log^2(1/\epsilon))}$.
This gives an improvement on the dependency on $k$ on the exponent, while having a worse dependency on $\epsilon$.
As in Laue and Matijevi\'{c}, we follow the framework introduced by Arora for Euclidean TSP.
Our key ingredients include exponential distance scaling and compression of dynamic programming state tables. 
\end{abstract}

\section{Introduction}

Given a set $S$ of $n$ points in $2$-dimensional Euclidean space, an
integer $k$, and a root node $r\in S$, we would like to find a
spanning tree with minimum cost rooted at $r$ such that any path from
$r$ to any point contains at most $k$ edges.  We refer to a spanning
trees satisfying this condition as a $k$-hop spanning tree.  This
problem has applications in network
design~\cite{carmi2019,carmi2015,carmi2018}, distributed system
design~\cite{Raymond89}, and wireless networks~\cite{Haenggi04}.

For any $\epsilon>0$, Laue and Matijevi\'{c}~\cite{laue2008} present a
polynomial-time approximation scheme (PTAS) for this problem on the
plane that runs in time $n^{O(k/\epsilon)}$ for finding a
$(1+\epsilon)$-approximate solution.  They follow the general
framework of random dissection by Arora~\cite{arora1996,arora1997} for
finding good approximate solutions for instances in Euclidean metric.
The dynamic programming structure of~\cite{laue2008} (reviewed in
Section~\ref{sect:dp-overview}) follows the approach from the PTAS for
the $k$-median problem by Arora, Raghavan, and Rao~\cite{arora1998}.

In this work, we give a novel trade-off between the hop bound $k$ and
the approximation requirement $\epsilon$.  Namely, we reduce the
dependency on $k$ on the {\em exponent} of the running time from $k$
down to $\log k$, while suffering a factor of
$(1/\epsilon)\log^2(1/\epsilon)$ increment.  This might not be of
an important concern in practice, but we would like to note that in many
other problems the hop bound can be seen as ``hard'' constraints;
therefore, (doubly) exponentially improvements on the dependency might
indicate possibilities for further improvements.

When the points are on a metric and $k=2$, Alfandari and
Paschos~\cite{AlfandariP99} show that the problem is MAX-SNP-hard;
thus it is unlikely to have a PTAS.

When points are randomly distributed in the $d$-dimensional Euclidean
space, Clementi~{\em et al.}~\cite{ClementiDIMLRS05} prove a
lowerbound on the cost of $k$-hop MST and show that a
divide-and-conquer heuristic finds a solution matching the lowerbound.

For general metrics, Althaus, Funke, Har-Peled, K\"{o}nemann, Ramos,
and Skutella~\cite{althaus2005} present an $O(\log n)$-approximation
algorithm based on low distortion tree embeddings.  Kantor and
Peleg~\cite{kantor2006} present a constant factor approximation
algorithm that runs in time $1.52\cdot 9^{k-2}$.

The problem can be generalized to the $k$-hop Steiner tree problem by
allowing the set of points required to be connected to $r$ to be
$X\subseteq S$.  B{\"o}hm~{\em et al.}~\cite{BohmHMNS20} gives an
$n^{O(k)}$ exact algorithm for this problem when the points are from
the metrics induced by graphs of bounded tree width.

We review the PTAS framework in Section~\ref{sect:prelim} and review
the dynamic programming approach of~\cite{laue2008} in
Section~\ref{sect:dp-overview}.  Section~\ref{sect:reduce} discusses
how we change the dynamic programming table to reduce the dependency
on $k$.  We give full descriptions of the dynamic programming in
Section~\ref{sect:compute-dp} and its analysis in
Section~\ref{sect:analysis}.


\section{Preliminaries}
\label{sect:prelim}
We begin with the description of the bounding box and quadtree
dissection based on techniques of Arora~\cite{arora1996,arora1996}
which is also used by Laue and Matijevi\'{c}~\cite{laue2008}.

\subsection{Bounding box}
Let the bounding box be the smallest axis-aligned square inside which
the set of points lie.  As in previous work
(e.g.,~\cite{laue2008,arora1996}), we can scale and translate all the
points into the bounding box of side length
$L={O(\frac{n}{\epsilon})}$.  We can also move all the points into the
closest grid point, while increasing the cost of the $k$-hop MST by at
most $\epsilon$ fraction of the optimal cost.  To see this, note that
the cost of the optimal solution is at least $\Omega(n/\epsilon)$, the
increased cost for each edge in the grid-aligned solution is at most
$4$, incurring the cost at most $4n$ for the entire solution, which is
only an $\epsilon$ fraction of $\Omega(n/\epsilon)$.

\subsection{Quadtree dissection}

A {\em dissection} of the bounding box is a recursive partition of a
square into four equal and smaller squares by vertical and horizontal
lines.  We recurse until the square has unit length or it contains at
most one point.  We call each square obtained from the procedure a
{\em box} in the dissection.  This recursive partition forms a
quadtree, whose nodes are boxes, rooted at the node representing the
bounding box.  We assign levels to the boxes in dissection as the
level of their associated nodes in the quadtree. The root node has
level 0.  There are at most $O(L^2)$ nodes and its depth is $O(\log
L)=O(\log (n/\epsilon))$.  We say that vertical and horizontal lines
are of {\em level} $i$ if we recursively partition a square at level
$i$ to level $i+1$ by these lines.

For each box, we introduce the set of pre-specified points on its side
called {\em portals}.  It is hard to find a solution when each
solution can cross a box at any position.  Therefore, we place $m$
equally spaced portals on each side of a box, with a total of $4m$
portals per box, to enforce a solution to cross each box only
at these portals.  If every edge of a solution of $k$-hop MST crosses
the sides of each box in the dissection only at its portals, we say
that the solution is {\em portal respecting}.

To deal with an increased cost of the optimal portal-respecting $k$-hop
MST solution, we use randomized shift described as follows.  Let $a$,
$b$ be a positive integer, the $(a,b)$-shift dissection is defined by
shifting all the lines with $x$ and $y$-coordinate by $a$ and $b$
respectively, and then modulo with $L$.  In other words, a vertical
line with $x$-coordinate $X$ will move to $x$-coordinate $(X + a)
\bmod L$ and a horizontal line with $y$-coordinate $Y$ will move to
$y$-coordinate $(Y + b) \bmod L$.  The following is a key lemma
from Arora~\cite{arora1996}.

\begin{lemma}
Let $m = O(\log L/\epsilon) = O(\log(n/\epsilon)/\epsilon)$ and choose
two positive integers $a$ and $b$ at random such that $0 \leq a, b <
L$, with probability at least $1/2$, there is an optimal
portal-respecting solution with respect to the $(a, b)$-shift
dissection of cost at most $(1 + \epsilon)$ times the optimal $k$-hop
MST solution.
\label{lem:arora}
\end{lemma}

We, later on, focus only on finding good portal-respecting solutions.



\section{Dynamic Programming}
\label{sect:dp}

We apply the dynamic programming approach to find the portal
respecting solution that has approximation ratio $(1+\epsilon)$ as in
Laue and Matijevi\'{c}~\cite{laue2008}.  We first review the approach
used by~\cite{laue2008}, which is introduced by Arora, Raghavan, and
Rao~\cite{arora1998} in Section~\ref{sect:dp-overview}.  We give an
overview of our improvements in Section~\ref{sect:reduce} and provide
the description of the algorithm in Sections~\ref{sect:table} and~\ref{sect:compute-dp}.

\subsection{Review: the Tables of Laue and Matijevi\'{c}}
\label{sect:dp-overview}


Later on, to distinguish from the levels of boxes, we refer to the hop
distance of a point from the root in the $k$-MST solution as its {\em
  hop-level}.  Consider a box $B$ with side length $l$, we can find
the optimal portal respecting $k$-MST solution inside $B$ if we have
information on points outside of $B$.  Suppose we want to assign a
point $x\in B$ to hop-level $i$ in the tree.  To do so optimally, we
need to know the closest point $y$ with hop-level $i-1$.  This point
might be outside $B$; therefore, while we work on box $B$, we need to
specify this as a requirement for this particular solution.

The approach introduced by~\cite{arora1998} is to represent these
requirements approximately with two assignments $inside$ and $outside$
on the portals.  We would later ``guess'' the values of these
assignments (by enumerating all possible approximate values) when
working on box $B$.

\begin{figure}
\centering
\includegraphics[scale=0.5]{{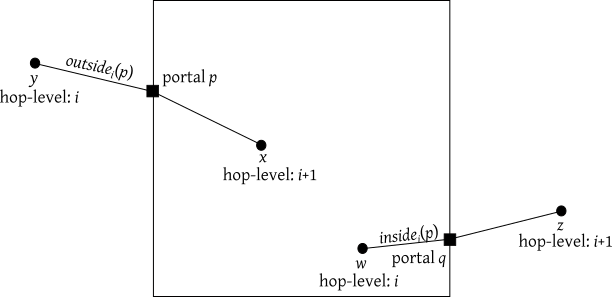}}
\caption{ Examples of $inside$ and $outside$ assignments:
  $outside_i(p)$ is the distance from portal $p$ to the closest point
  $y$ of hop-level $i$ outside the box, and $inside_i(q)$ is the
  distance from portal $q$ to the closest point $w$ of hop-level $i$
  inside the box.  Note that $outside_i(p)$ provides information for
  connecting $x$ inside the box, and $inside_i(q)$ provides
  information for connecting $z$ outside the box (in the higher level
  of the recursion).}
\label{fig:inside_outside}
\end{figure}

\begin{figure}
\centering
\includegraphics[scale=0.5]{{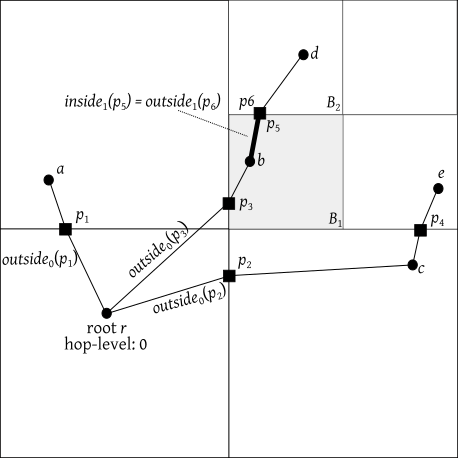}}
\caption{An example showing 2 levels of the recursive quadtree
  dissection.  Point $r$ is the root of the $k$-MST, points $a,b$, and
  $c$ are at hop-level $1$, and points $d$ and $e$ are at hop-level
  $2$.  Consider two boxes $B_1$ and $B_2$ at level $1$.  Inside
  $B_1$, the cost for point $b$, which connects to $r$ through portal
  $p_3$, can be computed using $outside_0(p_3)$.  Inside $B_2$, the
  cost for point $d$ of hop-level $2$, which connects to $b$ through
  portal $p_6$ can be computed using $outside_1(p_6)$.  Note that box
  $B_1$, which provides the connection for $d$, must guarantee the
  ``inside'' distance through portal $p_5$ (which is at the same
  location as portal $p_6$ of $B_2$).  Observe that, in this case,
  $inside_1(p_5)=outside_1(p_6)$. }
\label{fig:inside_outside_recur}
\end{figure}

To illustrate the idea, we would first start by describing the
assignments $outside$ and $inside$ in an ``ideal'' setting, where we
keep all distances exact.  For portal $p$ and hop-level $i$,
$outside_i(p)$ is the distance from $p$ to the closest point {\em
  outside} the box with hop-level $i$ and $inside_i(p)$ is the
distance from $p$ to the closest point {\em inside} the box with hop
level $i$.  One can view the $inside$ and $outside$ assignments as a
specification for a particular subproblem in box $B$ which includes a
provided ``outside'' condition in the $outside$ assignment and a
requirement from $inside$ assignment that has to be satisfied (by
providing points with appropriate hop-levels and portal-distances
``inside'' the box).  See Figure~\ref{fig:inside_outside} for an
illustration on $inside$ and $outside$ assignments and
Figure~\ref{fig:inside_outside_recur} for interactions between these
assignments between levels of recursion.

Since the actual values for $inside$ and $outside$ are reals, it is
not possible to enumerate their values during the dynamic programming
table evaluation.  We would instead deal with approximate values by
rounding these values up to be multiples of $l/m$.  Note that since
forcing the solution to be portal-respecting already introduces an
additional error of $l/m$ for each tree-edge going through a portal,
the analysis in Lemma~\ref{lem:arora} can be modified slightly to
account for an additional $l/m$ additive error for each tree edge in
the solution.

We would keep in assignments $outside$ and $inside$ approximate
distances.  Consider box $B$ whose side length is $l$.  For the
$inside_i$ assignments, we know that the distance between any two
points inside box $B$ is at most $2l$, and since we can tolerate $l/m$
errors, we only keep $inside_i(p) \in \{0, l/m, 2l/m, ..., 2l,
\infty\}$, for portal $p$, where $\infty$ represents the fact that no
nodes of hop-level $i$ are inside $B$.  Similarly, for the $outside$
assignments, we know that any two points in the bounding box can be at
most $2L$ apart; thus, we can have $outside_i(p) \in \{0, l/m, 2l/m,
..., 2L, \infty\}$, for portal $p$.

For each subproblem for box $B$, we want to find the minimum cost
solution satisfying the $inside$ and $outside$ assignments.

The table entry
\[
Table(B, inside_0, ..., inside_{k-1}, outside_0, ..., outside_{k-1})
\]
keeps the minimum cost solution for box $B$ that respects the
specified $inside$ and $outside$ assignments.

The key observation from~\cite{arora1998,laue2008} is that the
distance assignment between adjacent portals can differ by at most
$2l/m$; this implies that the number of possible assignments of
$inside_i$ is $2m\cdot3^{4m}$ per box.  With the same observation, the
number of possible assignments of $outside_i$ is $2Lm/l\cdot3^{4m}$ per
box. In total we have $4Lm^2\cdot3^{8mk}$ possible assignments we have
to ``guess'' during the dynamic programming evaluation per box.
Plugging in values for $L$ and $m$ yields that we have
$n^{O(k/\epsilon)}$ assignments for each box.  The running time
of~\cite{laue2008} depends essentially on this number of assignments,
because to compute all table entries for box $B$ with a particular
$inside$-$outside$ assignment, one have to go over all possible
assignments of 4 child boxes of $B$ and check their compatibility (in
$O(m^4)$ time); if the number of assignments per box is at most $T$,
the running time is bounded by $T\cdot T^4\cdot m^4 =
n^{O(k/\epsilon)}$.

\subsection{Reducing the table size}
\label{sect:reduce}

We describe two basic ideas for reducing the table size.  Let
$\delta=\delta_\epsilon$ be a constant depending on $\epsilon$, to be
defined later.  Our goal is to ensure that the multiplicative error
incurred in each subproblem is at most $1+c\cdot\delta$, for some
constant $c$.  Section~\ref{sect:error-bounds} analyzes the value for
$\delta$ and bounds the total error.

\subsubsection{Scaling the distances}

Consider box $B$ whose side length is $l$.  Consider a portal $p$ of
$B$.

Instead of using a fixed linear scale $l/m$, we use exponentially
increasing distance scale.  This is a standard technique; see, e.g.,
Kolliopoulos and Rao~\cite{kolliopoulos2007}.  For the $inside$
assignments, we choose distances from the set $\{0,l/m, (1+\delta)l/m,
(1+\delta)^2l/m, (1+\delta)^3l/m,\ldots\}$.  This reduces the number
of states for each portal from $O(m)$ to $O(\log_{1+\delta} m)$.  We
also use the same trick for $outside$, i.e., we allows the distances
from the set $\{0, l/m, (1+\delta)l/m, (1+\delta)^2l/m, \ldots, L\}$.
The number of possible distance values is $O(\log_{1+\delta} L/m)$.

Using exponentially increasing distance scales clearly reduces the
number of states, however, disparity between the actual distance when
merging subproblems incurs additional multiplicative error to our
solution.  We would bound this error in
Subsection~\ref{sect:error-bounds}.

We also slightly change how the distance assignments $inside$ and
$outside$ are represented in the box.  For each portal, instead of
keeping the distance for each hop-level $i$, we keep for each distance
scale the minimum hop-level $i$ of nodes within that distance.
Formally, let $\gamma_0=0, \gamma_1=(1+\delta)^0l/m, $ and, in
general, $\gamma_j=(1+\delta)^{j-1}l/m,$ be distance level
thresholds.  When consider distance level $j$, for $j\geq 0$, for
portal $p$, we look for the minimum hop-level of a node with distance
at most $\gamma_j$.  Therefore, for each distance level, there are
$k^{O(m)}$ possible value assignments.  Considering all distance levels, there are
$k^{O(m\log_{1+\delta} L)}$ value assignments, which are too many.  We need
another idea to reduce this number.

\subsubsection{Granularity for hop changes}
\label{sect:granularity}

We can further reduce the number of possible values for adjacent portals.  Since adjacent
portals in $B$ are $l/m$ apart, if we can tolerate larger error, we
can let portals share the same closest point.  More specifically,
since we can tolerate a factor of $1+\delta$ multiplicative error, if
the actual distance is at least $(l/m)(1/\delta)$, additional error of
$l/m$ is acceptable.

Consider distance level $j'$ where
\[
(1+\delta)^{j'}(l/m)\geq \frac{2}{\delta}(l/m).
\]
In that level, we can let two adjacent portals $p$ and $p'$ share the
closest point.  This reduces the number of variables at this level by
half.  Suppose that at distance level 1 there are $k^{O(m)}\leq
k^{cm}$ value assignments for some constant $c$, in this level $j'$, there will
be $k^{cm/2}$ value assignments.  More over at level $j''$ where
$(1+\delta)^{j''}\geq 4/\delta$, we can further combine portals and
there will be only $k^{cm/4}$ value assignments.  Let $\beta$ be the smallest
integer such that
\[
(1+\delta)^{\beta}\geq \frac{2}{\delta},
\]
i.e.,
\[
\beta\geq \frac{\log(2/\delta)}{\log(1+\delta)}.
\]
Thus, the number of possible value assignments decreases at every $\beta$
distance levels.  This implies that the number of assignments is
\[
(k^{cm})^{\beta} \cdot
(k^{cm/2})^{\beta} \cdot
(k^{cm/4})^{\beta} \cdot\ \ \cdots,
\]
which is $k^{O(m\beta)} = (n/\epsilon)^{O(\beta\log k/\epsilon)}$,
since $m=O(\log L/\epsilon)$.  We would work out the value of $\beta$
later in Section~\ref{sect:running-time}.

\subsection{The Table}
\label{sect:table}

From the discussion in Section~\ref{sect:reduce}, we formally describe
our compressed table.  Consider box $B$ with side length $l$ whose
portals are $p_1,p_2,\ldots,p_{4m}$.  Recall that
$\gamma_j=(1+\delta)^{j-1}l/m$ is the $j$-th distance level threshold
for $B$.  Let $\alpha_i=1 + \log_{1+\delta} m$ and $\alpha_o = 1 +
\log_{1+\delta} L$ be the number of distance levels for $inside$ and
$outside$ assignments.  For a set of portals $p_s,\ldots,p_t$ of $B$,
let

\begin{itemize}
\item $ilevel_{j}(s,t)$ for $j=0,1,\ldots,\alpha_i$ be the
  minimum hop-level $i$ satisfying $inside_i(p_d)\leq\gamma_j$ for
  some $s\leq d\leq t$,
\item $olevel_{j}(s,t)$ for $j=0,1,\ldots,\alpha_o$ be the
  miniumum hop-level $i$ satisfying $outside_i(p_d)\leq\gamma_j$ for
  some $s\leq d\leq t$.
\end{itemize}

Also recall $\beta=\lceil \log_{1+\delta}(2/\delta) \rceil$.  Instead
of distance assignments $inside$ and $outside$ we keep the following
assignments.  For levels $j=0,1,\ldots,\beta-1$ and $1\leq d\leq 4m$,
assignments $ilevel_j(d,d)$ and $olevel_j(d,d)$.  For levels
$j=\beta,\beta+1,\ldots,2\beta-1$ and $1\leq d\leq 2m$, assignments
$ilevel_j(2d-1,2d)$ and $olevel_j(2d-1,2d)$.  In general, for
$f=1,2,\ldots,$ for levels $j=f\cdot\Delta, f\cdot\beta+1, \ldots,
(f+1)\cdot\beta-1$, and $1\leq d\leq 4m/2^f$, assignments
\[
ilevel_j(2^f (d-1)+1,2^f d),
\]
and
\[
olevel_j(2^f (d-1)+1,2^f d).
\]

We refer to this set of assignments as a compressed representation.
As noted in Subsection~\ref{sect:granularity}, the number of
assignment variables in a box for the first $\beta$ levels is
$O(m\beta)=O(m\log_{1+\delta} (2/\delta))$.  The number decreases
exponentially for every $\beta$ levels; thus there are at most
$O(m\log_{1+\delta} (2/\delta))$ variables, implying the total number
of $k^{O(m\log_{1+\delta} (2/\delta))}$ assignments per box.

From distance assignments $inside_i$ and $outside_i$ described
in~\cite{laue2008}, one can obtain the compressed representation in
polynomial time.  Furthermore, from a compressed representation, the
original assignments $inside$ and $outside$ can also be recovered in
polynomial time with, possibly, missing values.  This issue can be
dealt with by filling them up with $\infty$.  With this
polynomial-time transformation, to compute the dynamic programming
table, we can use the algorithm of Laue and
Matijev\'{i}c~\cite{laue2008} with only slight modification.

\subsection{Computing The Table}
\label{sect:compute-dp}

As mentioned in the previous section, we can use the dynamic
programming algorithm of Laue and Matijevi\'{c}~\cite{laue2008} to
solve the problem.  Here we describe a slight modification of their
algorithm (to handle missing values) in detail for completeness.
Note that the time bound for their algorithm essentially depends on
the total number of states, not the merging procedure.

\subsubsection{Base case}
Consider box $B$ which is the leaf of the recursive quadtree
dissection. Let $dist(u, v)$ be the Euclidean distance between points
$u$ and $v$. As in~\cite{laue2008}, we consider two base cases.

{\bf Case 1:} Box $B$ contains root $r$.  In this case, the box may
contain other points, but they all lie at the same position as $r$.

For any non-root node inside box $B$, we create an edge between them
and root $r$ (with cost $0$).  We set a table entry of box $B$ with
cost $0$ if for each portal $p$,
\begin{enumerate}
\item $outside_i(p) = \infty$, for $0 \leq i \leq k-1$,
\item $inside_i(p) = \infty$, for $1 \leq i \leq k-1$, and
\item $dist(r, p) \leq inside_0(p)$.
\end{enumerate}
  
For all other entries of box $B$, we set their costs to be $\infty$.

{\bf Case 2:} All points inside box $B$ are at the same grid point $u$
and box $B$ does not contain root $r$.

Consider each possible entry for $B$.  We first ensure that the
$inside$ assignments can be satisfied at hop-level $i$, i.e., that
$dist(u,p)\leq inside_i(p)$ for each portal $p$ and
$inside_{i'}(p)=\infty$ for all $i\neq i'$.  If this condition is not
satisfied, we set the cost of this entry to $\infty$.

We have to connect a point at $u$ to some point outside $B$ at
hop-level $i-1$.  As mentioned in Laue and
Matijev\'{i}c~\cite{laue2008}, there are two possible cases: (1)
either each point in $B$ is connected with its own tree edge through a
portal $p$ (where each point pays a separate connection cost) or (2)
only one point $q$ is connected with its own tree edge through a
portal (paying the cost) while the others connect through $q$ (paying
$0$ cost) with one additional hop-level.  Note that the latter case is
always cheaper, but it might violate the hop constraint; therefore only
applicable when the hop-level $i<k$.

More precisely, when hop-level $i<k$, we set the cost for this entry
to be
\[
\min_{p'} outside_{i-1}(p') + dist(u,p'),
\]
where $p'$ ranges over all portals of $B$.  Otherwise, when $i=k$, we
set the cost to be
\[
|B|\cdot \left(\min_{p'} outside_{i-1}(p') + dist(u,p')\right),
\]
where $|B|$ is the number of points in box $B$ and $p'$ ranges over
all portals of $B$.

\subsubsection{Merging process}

Let $B$ be the box in dessection with children $B_1,B_2,B_3,$ and
$B_4$.  For each assignment of $inside$ and $outside$ of $B$, we
consider all assignments $inside^{(j)}$ and $outside^{(j)}$ of its
children $B_j$, for $1\leq j\leq 4$, to compute $B$'s entry of the
table.  We have to check (1) if some $B_j$ provides the required
$inside$, and (2) if the children satisfy their own $outside^{(j)}$
requirements with some $inside^{(j')}$ or $B$ appropriately propagates
the $outside$ requirements.  If all conditions are satified, we set
the cost of $B$'s entry to be the minimum of the sum of the children
costs.  Otherwise, we set the cost to be $\infty$.

\section{Analysis}
\label{sect:analysis}

\subsection{Error bounds}
\label{sect:error-bounds}

During the merging process, for each level, we can incur additional multiplicative factor of $(1+\delta)$.  Since there are $O(\log (n/\epsilon))\leq c\log (n/\epsilon)$ recursive levels, if we let $\delta$ to be such that $(1+\delta)^{c\log (n/\epsilon)}\leq (1+\epsilon)$, we can ensure that the error is not too large.  However, this implies that $\delta\leq \epsilon/\log (n/\epsilon)$, which is too small, and we have to take an additional $\log n$ on the exponent, resulting in a quasi-polynomial time algorithm.

We shall give a better analysis on the error so that we only need $\delta = O(\epsilon/\log (1/\epsilon))$.  Consider an edge $(u,v)$ in an optimal solution whose length is $\ell$.  From the randomized dissection, we know that the expected portal respecting length of this edge is at most $(1+\epsilon)\ell=\ell + \epsilon\ell$.  We analyze additional errors due to exponential distance level scaling.

Let box $B$ be the box in the dissection that contain both $u$ and $v$.  Clearly, all relevant boxes are those in $B$ that contain $u$ or $v$.  First note that boxes containing $u$ or $v$ whose side lengths are at most $\epsilon\ell$ contribute to at most $O(\epsilon\ell)$ additive errors; thus we only consider boxes containing $u$ and $v$ whose side lengths are at least $\epsilon\ell$.

We first deal with relevant boxes containing $u$ in $B$ whose side lengths are at most $4\cdot\ell$.  The number of recursive levels is $\log \left(\frac{4\ell}{\epsilon\ell}\right)=O(\log (1/\epsilon))$.  Each level the multiplicative error can be at most $(1+\delta)$.  
For relevant boxes in $B$ whose sides are larger than $4\ell$, we claim that $(u,v)$ can pass through at most 2 of them, incurring at most another $(1+\delta)^2$ factor.  To see this, suppose that $(u,v)$ passes through one vertical side of box $B'$ at level $j$ whose side length is larger than $4\ell$.  Since $(u,v)$'s length is $\ell$, it cannot reach another vertical line at level $j-1$.  The same argument applies to horizontal sides.

Thus, to ensure that the accumulative multiplicative error over all these levels is at most $(1+\epsilon)$, we need $\delta$ to satisfy
\[
(1+\delta)^{\log (1/\epsilon)+2}\leq 1+\epsilon.
\]
Note that since $1+x\leq e^x$, the following inequality suffices
\[
(1+\delta)^{\log (1/\epsilon)+2}\leq e^{\delta(\log (1/\epsilon)+2)}\leq 1+\epsilon;
\]
thus, by taking log, we need to ensure that
$\delta=O\left(\frac{\log(1+\epsilon)}{\log (1/\epsilon)}\right)
=O\left(\frac{\epsilon}{\log(1/\epsilon)}\right)$, because $\epsilon/2
\leq \log(1+\epsilon)$, for $\epsilon\leq 1$.

\subsection{Running time}
\label{sect:running-time}

Again note that the running time mainly depends on the number of
assignments per box.  From Section~\ref{sect:granularity}, we show
that the number of assignments for each box is
\[
k^{O(m\beta)} = (n/\epsilon)^{O(\beta\log k/\epsilon)}.
\]
We are left to bound $\beta$, the number of distance levels needed before
we can combine portals.  Recall that we need
\[
\beta \geq \log(2/\delta)/\log(1+\delta).
\]
Since $\delta\leq 1$, we know that
$\delta/2\leq\log(1+\delta)$. Combining this with the bound from the
previous section, we can let
\[
\beta
= O\left(
\frac{\log(\log(1/\epsilon)/\epsilon)}
     {\epsilon/\log(1/\epsilon)}
     \right)
     = O\left(\frac{\log(1/\epsilon)\cdot\log(\log(1/\epsilon)/\epsilon)}
{\epsilon}
\right)
= O\left(\frac{\log^2(1/\epsilon)}{\epsilon}\right)
\]
since $\log(\log(1/\epsilon)/\epsilon)=\log\log(1/\epsilon) + \log(1/\epsilon)=O(\log(1/\epsilon))$.

Let $T$ be the maximum number of assignments per box.  Clearly since the
merging process takes at most $O(T^5)$-time per box, we only need to
analyze the number of assignments.  From the discussion in
Section~\ref{sect:table}, the number of assignments $T$ is
\begin{align*}
k^{O(\beta\cdot m)}
&= k^{O(\beta\cdot\log (n/\epsilon)/\epsilon)} \\
&= (n/\epsilon)^{O(\beta\cdot \log k/\epsilon)}\\
&= (n/\epsilon)^{O((\log^2(1/\epsilon)/\epsilon)\cdot(\log k/\epsilon))}\\
&= (n/\epsilon)^{O(\log k\cdot(\log(1/\epsilon)/\epsilon)^2)}
\end{align*}
Thus, we have the running time bound as claimed.


\bibliographystyle{plain}
\bibliography{references}




\end{document}